\documentclass{aa}  
\usepackage{graphicx}
\usepackage[authoryear]{natbib}
\bibpunct{(}{)}{;}{a}{}{,}
\usepackage{txfonts}
\usepackage{longtable}
\begin{document}
\title{Analysis of the stellar population in the central area of the HII 
region Sh 2-284\thanks{Based on observations made with the 
Nordic Optical Telescope, 
operated on the island of La Palma jointly by Denmark, Finland, Iceland,
Norway, and Sweden, in the Spanish Observatorio del Roque de los
Muchachos of the Instituto de Astrofisica de Canarias.}}

   \author{Antonio J. Delgado\inst{1}, Anlaug A. Djupvik\inst{2}
          \and
           Emilio J. Alfaro\inst{1}
          }

   \offprints{A.J. Delgado}

   \institute{Instituto de Astrof\'\i sica de Andaluc\'\i a, CSIC, \\
    Apdo 3004, 18080-Granada, Spain\\
              \email{delgado@iaa.es},\email{emilio@iaa.es}
         \and
Nordic Optical Telescope,
Apado 474, 38700 Santa Cruz de La Palma, Spain \\
             \email{amanda@not.iac.es}
             }

   \date{Received April 15,2008; accepted April 31, 2008}

 
  \abstract 
{There is a lack of state-of-the-art information on very young open 
clusters, with implications for determining the structure of the Galaxy.}  
{Our main objective is to study the timing and location of the star formation 
processes which yielded the generation of the giant HII region Sh 2-284. This 
includes the determination of different physical variables of the stars, such 
as distance, reddening, age and evolutionary stage, including
pre-main-sequence (PMS) stars.} 
{The analysis is based on $UBVR_CI_C$ CCD measurements of a field of 
6.5'$\times$6.5' containing the cluster, and JHK$_s$ photometry in the 
3.5'$\times$3.5' subfield, centered in the apparent higher condensation. 
The determination of cluster distance, reddening and age is carried out 
through comparison with ZAMS, post-MS and PMS isochrones. The reference 
lines used are obtained from theoretical post-MS and PMS isochrones from 
the Geneva and Yale groups, for metalicity Z=0.004, in agreement with the 
spectroscopic metallicity determination published for several cluster members.}
{The results amount to E(B-V)=0.78$\pm$0.02, DM=12.8$\pm$0.2 (3.6~kpc), 
LogAge(yr)=6.51$\pm$0.07 (3.2 Myr). The distance result critically 
depends on the use of low metallicity ZAMS and isochrones. A PMS member 
sequence is proposed, with an age of LogAge(yr)=6.7$\pm$0.2 (4.7 Myr) 
which is therefore coeval within the errors with the post-MS cluster age. 
The mass function for this population in the mass range above 1.3-3.5 
M$_\odot$ is well fitted by a Salpeter mass function. The presence of a 
different star generation in the cluster with a distinctly older age, 
around 40 Myr, is suggested. On the other hand, the NIR photometry 
results indicate a large number of sources with (H-K$_s$) excess, 
practically distinct from the optical PMS candidate members. }
  {The analysis of our deep $UBVRIJHK$ photometry of Dolidze 25 therefore 
reveals a young cluster with coeval MS and PMS populations of age 3.2-5 Myr. 
In addition, a distinctly older cluster member population of age 40 Myr is 
suggested. The distance determined for the cluster from quantitative fits to 
ZAMS and isochrones is distinctly lower than previously published values. 
This result originates in the consistent use of low metallicity
models for ZAMS fitting, applying published metallicity values
for  the cluster}

\authorrunning{A.J. Delgado, A.A. Dujpvik \& E.J. Alfaro} 
\titlerunning{The central area of the H{\sc ii} region Sh 2-284}

   \maketitle
%

\section{Introduction}

The open cluster Dolidze~25 (06h 45m 06s, +00$^{\circ}$ 18' 00'', Epoch 2000) 
is one of the targets in our long term project on the search for and 
characterization of PMS members in young open clusters \citep[and 
references therein, DAY-I in the following]{del07}. The cluster was 
included in their series of photoelectric studies by \citet{mof75} 
and \citet{mof79}. A UBV CCD photometric study was more recently 
published \citep{tur93}.

In addition to the general objectives of this project (DAY-I), 
Dolidze~25 presents in principle two additional features of interest. 
It is located in the Galactic anticenter direction, and has been 
claimed to have a metallicity distinctly lower than the solar value 
\citep{len90,fit92}. The cluster is therefore of special interest in 
our project, because of the expected finding of PMS stars with low 
metallicity and possibly at large Galactocentric distances. 

The low metallicity of the cluster would contradict the value expected from 
the published values of the Galactic metallicity gradient. Any of the published values obtained from B-type stars, between -0.05 to -0.07~dex/kpc \citep{fu09}, would require too large Galactocentric distances for our cluster to reach the \cite{len90} metallicity value. On the other hand, recently published results \citep{prza08,prz08}, with an improved method of spectroscopic analysis of B-type stars, suggest a clearly tighter result for abundances of B-type stars in the solar neighbourhood than those found in previous works. The accurate abundances could be considered to define the zero point of B-type stars metallicity at the solar Galactocentric distance, which would imply necessary corrections of previous values towards higher metallicities. We note however, that local metallicity variations with respect to an average gradient have been reported several times \citep{fu09,ped09}, and the possible presence of discontinuities and slope changes, especially for large Galactocentric distances are still open.
In this context, the analysis of cluster 
parameters, and the eventual improvement of their values is of importance. 
The cluster was selected as target for one of the 
COROT\footnote{http://earth-sciences.cnes.fr/COROT/} Additional Programs 
\citep{rip06}. These authors performed VIMOS observations at VLT 
which revealed the presence of a large number of emission line objects 
in the region (partial report 
ftp://ftp.na.astro.it/pub/astrows06/presetazione-FCUSANO.ppt at 
http://www.na.astro.it/; http://earth-sciences.cnes.fr/COROT/A-corot-week.htm).

Finally, the cluster is located in the center of the giant H-alpha bubble 
which encompasses the HII region Sh 2-284 (shortened to S284 in the 
following). 
Thus the analysis of the stellar content of this field could help to better 
understand the connection between the star formation processes and the 
physics of the ionized gas. The S284 region has been the object of a 
mid-infrared study with Spitzer IRAC bands \citep{pug07}. The analysis of 
these observations \citep[P09 in the following]{pug09} includes Dolidze~25, 
and addresses the properties of massive star formation in the region.

\defcitealias{del07}{DAY-I}
\defcitealias{pug09}{P09}

In the present study we report UBVRIJHK observations centered in the 
Dolidze~25 cluster. In the next section we give an overview of the region. 
In Sect.~\ref{obs} we describe the different data sets, with reduction 
procedures. Section~\ref{ubvri-results} presents the optical results and 
a discussion of the determination of cluster parameters and cluster 
membership. Section~\ref{jhk-results} presents the near-IR results and 
IR-excess sources. Section~\ref{disc} contains the discussion of the 
results, and finally the last section resumes the main conclusions.

\section{Overview of the region}
\label{overview}

\begin{figure}
\centering
\includegraphics[angle=0,width=9cm]{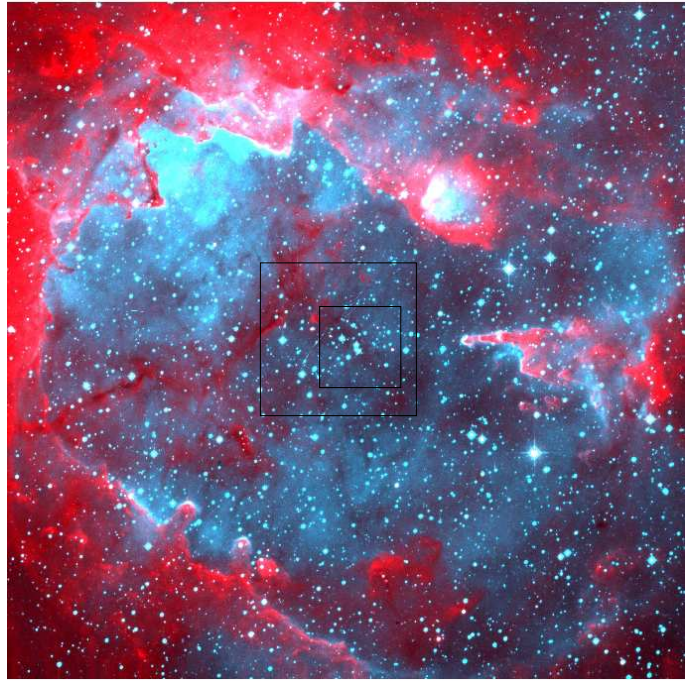}
\caption{Colour composite of 30' $\times$ 30' from AAO/UKST-H$\alpha$ 
survey (blue) and Spitzer/IRAC/8$\mu$m image (red). North up, east left. 
Dolidze~25 is located in the center of the bubble. 
In this paper we focus on the central area outlined by the larger and 
smaller squares representing the coverage of deep $UBVRI$ and $JHK_S$ 
photometry, respectively. 
}
\label{fig-UKST}
\end{figure}

Dolidze~25 is located in the  apparent center of an IR dust bubble 
\citepalias{pug09}.  
In this article we focus on a photometric study in UBVRIJHK of the central 
Dolidze~25 cluster, and its inmediate neighbourhood. 
In Fig.~\ref{fig-UKST} we show the emission from the 
8.0 $\mu$m Spitzer/IRAC band (red) obtained from the Spitzer archive and 
the H$\alpha$ emission (blue) from the AAO/UKST-H$\alpha$ survey 
(http://www.roe.ac.uk/ifa/wfau/halpha/) in a 30' $\times$ 30' area 
centred on the cluster. The fields covered by the $UBVRI$-ALFOSC and the 
$JHK_S$-NOTCam images are outlined by a larger and a smaller square, 
respectively. The projected circular structure is formed by several 
knots and clouds of H$\alpha$ emitting material, thoroughly discussed in 
\citetalias{pug09}. The image suggests a radial connection of several 
features to the cluster center. 
We recall here the presence of several arc-like structures, seen to the north 
and south, and two particularly interesting more defined features. To the 
north-east we see structures resembling broad arrow peaks, and to the west, a 
structure which very closely resembles a forefinger directly pointing to the 
central cluster, resembling the so-called elephant trunk features observed in 
some nearby star forming regions. All these structures seem to be 
geometrically related to the center, the location where Dolidze~25 lies.


\begin{table*}
\caption{UBVRIJHK photometry of 1673 sources in Do25, sorted on ascending
RA. ID is our id number, IDW is the WEBDA id number, and 2MASS is the
name of the cross-correlated 2MASS source. Positions are given in decimal
degrees and have epoch 2000. One source is listed as an example.}

\begin{tabular}{lllccrrrrrrrrrrrrrrrrcc}
\hline
ID & IDW & 2MASS & RA & DEC & U-B & $\sigma_{UB}$ & B-V & $\sigma_{BV}$
& V & $\sigma_{V}$  \\ 
   & V-R & $\sigma_{VR}$ & V-I & $\sigma_{VI}$ & J & $\sigma_{J}$ \\ 
   & H & $\sigma_{H}$ & K & $\sigma_{K}$ & MEM & IREX \\
\hline
1053 &  16 & 06450210+0013165 &   101.2587814 & 0.22125500 & -0.330 & 0.004 &
0.601 &  0.004 & 13.052 & 0.003 \\ 
   & 0.382 & 0.008 & 0.916 & 0.008 & 12.014 &  0.017 \\ 
   & 11.811 & 0.019 & 11.730 &  0.020 &  1 & 0 \\

\hline
\end{tabular}
\label{tab-1}
\end{table*}


\section{Observations and models}
\label{obs}

\subsection{$UBVRI$ photometry}

Deep $UBVRI$ imaging was made with ALFOSC at the 2.6m Nordic Optical 
Telescope (NOT) during available slots in the nights 25, 26 and 28 of 
December 2006, and the field was calibrated with photometric standards 
on the photometric night 21 of February 2007. The detector was a 
13.5$\mu$m $\times$ 2024 $\times$ 2024 pixel back-illuminated E2V CCD, 
covering a FOV of 6.5' with a pixel scale of 0.19''. 

The image reduction was carried out with adequate routines in the 
IRAF package. In every filter both short and long exposure times were 
secured to cover an as wide as possible brightness range. The long 
exposure times were 2700, 1800, 1800, 1800, and 3~$\times$~900 seconds, 
for UBVRI, respectively.  
The final instrumental PSF magnitudes were obtained from the respective 
frames, computing average values for the stars in common in both 
exposures. These final magnitudes are then best calibrated by direct 
correlation to published $UBV$ observations. In the present case, we 
used the published CCD $UBV$ colors \citep{tur93} better than the available 
photoelectric magnitudes by \citet{mof75} which coincide only 
partially with our field. 
The $(V-R)$ 
and $(V-I)$ color indices were calibrated with the catalogued values 
for 8 standard stars observed in the \citet{lan92} region Rubin149.
The uncertainties of our standard indices are calculated as the root 
mean squared deviations of the averaged O-C values for the different 
color indices. They amount to 0.05 mag in $V$, $(U-B)$ and $(B-V)$, 
0.01 mag in $(V-R)$ and 0.02 mag in $(V-I)$. 

\subsection{$JHK$ photometry}
\label{jhk-obs}

Deep near-IR imaging was obtained 
with NOTCam\footnote{See URL http://www.not.iac.es:/instruments/notcam/ 
for details on NOTCam.} at the NOT on the 14 of January 2006. The 
detector, a $1024 \times 1024 \times 18 \mu m$ Hawaii array, the first 
Science Grade Array for NOTCam (SWIR2), was available in the period 
October 2005 to May 2006. Its gain was 2.2~$e^-$/adu and the readout 
noise 15~$e^-$. The detector was found 
to be linear to within 1\% up to 30000 adu, with saturation starting 
at 54000 adu. About 2\% of the pixels were bad, but mainly along the 
edges of the detector. With this array the zeropoints for NOTCam were 
24.02, 23.97, and 23.33 mag (for 1 e$^-$/s, and Vega magnitudes) in 
the bands $J$, $H$, and $K_S$, respectively. 
We used the wide field camera (0.234''/pix, fov=4') and the broad band 
filters $J$, $H$ and $K_S$  with central wavelengths at $\lambda\lambda$ 
1.247, 1.632, and 2.140 $\mu$m, respectively, to map the central part 
of the cluster in a 3~$\times$~3 dither pattern. 
Exposing 6 $\times$ 6 seconds $\times$ 9 pointings $\times$ 3 
repetitions gave 162 images and a total on-source integration time of 
972 seconds per filter. 

We used the IRAF package and a set of our own scripts 
to mask bad pixels, subtract the sky background and flatfield the images 
using differential twilight flats taken on the same night. 
After image registration, combination and trimming, the final image size 
is reduced to 3.5' $\times$ 3.5' (see Fig~\ref{fig-jhkimage}). Point
sources were extracted from the final image of each filter using 
{\tt daofind}, and aperture photometry was made with an aperture radius 
of the order of the FWHM of the PSF. The flux loss in the PSF wings was 
corrected for by evaluating an aperture correction for each band on a 
few bright and isolated stars (with errors $<$ 0.014 mag for all bands). 
Errors related to flat-fielding 
were found to be 0.013, 0.012 and 0.015 mag in $JHK$, respectively. 
The above errors are added in quadrature to the error outputs from the 
IRAF task {\tt phot} to give conservative measurement errors. We reach 
down to $JHK$ magnitudes of 21.2, 20.0, and 19.5, respectively, with 
errors $<$ 0.30 mag. The median errors in $JHK$ are 0.047, 
0.040, and 0.044 mag, respectively, for the whole sample. The limiting 
magnitudes when requiring the total measurement errors to be $<$ 10\%, 
are 19.7 mag, 18.9 mag, and 18.6 mag in $J$, $H$, and $K$, respectively. 
We recall that the observations consist of multiple short integrations, 
obtained in worse than average seeing conditions (FWHM = 0.7'' in $K$). 
We estimate completeness down to at least $J$ = 18 mag, $H$ = 17.5 mag, 
and $K$ = 17 mag.

The 2MASS point source catalogue \citep{skr06} was used to calibrate 
both astrometry and photometry, and in regions not covered by NOTCam we 
used the 2MASS fluxes. We identified 85 common sources and made 
a plate solution of the NOTCam images with an rms in RA and DEC of 0.15" 
and 0.14", respectively. All sources were cross-correlated with the 
2MASS catalogue using the task {\tt tmatch} with a search radius of 
0.5'' ($>$ 3 times the rms). The fluxes were calibrated using the 19 
stars that satisfy the following three criteria: 1) K $<$ 14 mag, 
2) have 2MASS ABCD flags, and 3) not being YSO candidates, neither 
IR-excess sources (see Sect.\ref{jhk-results}) nor optically determined 
PMS stars (see Sect.\ref{pms-mem}). The difference magnitudes of these 
19 stars (NOTCam - 2MASS) were found to have a standard deviation of 
0.06~mag for $J$ and $K$ and 0.04~mag for $H$. 
Three bright stars (WEBDA id 15, 17, and 23) have fluxes exceeding the 
1\% linear range of the detector, and we therefore do not include their 
NOTCam $JHK$ measurements in the table of photometric results. 


\subsection{On-line table of $UBVRIJHK$ photometry.}

In Table~\ref{tab-1} (only available on-line) we list our photometric 
UBVRIJHK results in the observed fields. Table~\ref{tab-1} contains 1673 
sources, sorted on ascending RA, and
the columns give: Our ID number, WEBDA ID number, 2MASS name, RA, DEC, 
both in degrees,
U-B, $\sigma_{U-B}$, B-V, $\sigma_{B-V}$, V, $\sigma_{V}$, V-R,$\sigma_{V-R}$, 
V-I, $\sigma_{V-I}$, J, $\sigma_{J}$, H, $\sigma_{H}$, $K_S$, $\sigma_{Ks}$,
MEM, IREX. The column MEM shows membership (1 = young MS and post-MS 
member, 2 = old MS and post-MS member, 3 = PMS member according to 3 CM 
diagrams, 4 = PMS member according to 4 CM diagrams); see 
Sect.~\ref{ubvri-results} for details. The column IREX 
defines the possible presence of IR excess (0 = no IR excess, 1 = 
probable IR excess, 2 = bona fide IR excess), for details see 
Sect.~\ref{jhk-results}.

\subsection{Low metallicity ZAMS and isochrones} 

The low metallicity of Dolidze~25 makes it necessary to consider the 
consequent changes of the photometric indices. Mainly the value of the 
absolute visual (M$_V$) magnitude is affected by changes in metallicity, 
and its variation has been formulated as a linear dependence on metallicity 
for a given $(B-V)$ color \citep{van89}. For the analysis of Dolidze~25, 
in absence of empirical or semi-empirical ZAMS for metallicities lower 
than solar, we use theoretical ZAMS and isochrones computed for the 
corresponding metallicity, and transformed to the observational plane.

In addition to the differences between models, originating in the 
consideration and treatment of the various physical processes involved 
in the evolutionary calculations, it is this transformation to 
observable colors the factor that introduces the largest differences 
between the different models \citep{lej01}. To maximize 
the internal consistency of our results, we adopt reference lines 
transformed with the same calibration formulae to the observed colors. 
We use the post-MS Geneva isochrones \citep{lej01}\footnote{http://vizier.cfa.harvard.edu/viz-bin/VizieR?-source=VI/102}, 
and the Yale PMS isochrones \citep{yi01} in the 
Y2\footnote{http://csaweb.yonsei.ac.kr/~kim/y2solarmixture.htm} model 
set, namely those translated to the observational plane with the color 
transformation by \citet{lej98}. In both cases, the metallicity tracks 
for Z=0.004, as given for the cluster by \cite{len90}, are used. 
Neither the Geneva nor the Y2 models contain a nominal ZAMS line. We 
use here a composite curve, obtained from the Geneva isochrone of  
age 1000 yr for colors bluer than $(B-V)=0$, and the Y2 isochrone of 
age 40 Myr for redder $(B-V)$ color. We refer to this ZAMS line as 
ZAMS-Z004 in the considerations below.
Finally, the Y2 models show that the Color-Color (CC) relation for PMS 
isochrones deviates from the ZAMS line, specially for 
spectral types later than B0, a deviation which is larger the younger 
the isochrone age. The particular $(U-B),(B-V)$ relation for each 
isochrone is then used to compute color excess values, used afterwards 
in the computation of visual absorption and corresponding distance 
modulus with respect to this PMS isochrone.

%
%
\section{Results of UBVRI imaging}
\label{ubvri-results}

\subsection{Reddening law towards Dolidze 25}
\label{reddening}

As explained in \citetalias{del07}, the 
determination of cluster parameters starts with the selection of bona-fide 
unevolved MS members, on the basis of the photometric colors, with the 
help of membership assignments by other authors, and the eventual use of 
spectroscopic observations. The application of the method requires an 
assumption of the reddening slope $\alpha \equiv E(U-B)/E(B-V)$ used to shift 
the stars in the Color-Color (CC) $(U-B),(B-V)$ diagram. Once a 
reliable sample of non-evolved MS members is selected, they can be used to 
estimate the absorption coefficient $R_V=A_V/E(B-V)$, appropiate for the 
extinction law in the direction of the observed field, and to be applied in 
the calculation of distance modulus. 

\cite{tur93} derive $\alpha$=0.8 from a simultaneous fit of solar metallicity 
isochrones to their selected members in the $V,(U-B)$ and $V,(B-V)$ CM 
diagrams, with distance modulus, age, $E(U-B)$, and $E(B-V)$ as fitting 
parameters.


\begin{figure}
\centering
\includegraphics[angle=-90,width=8cm]{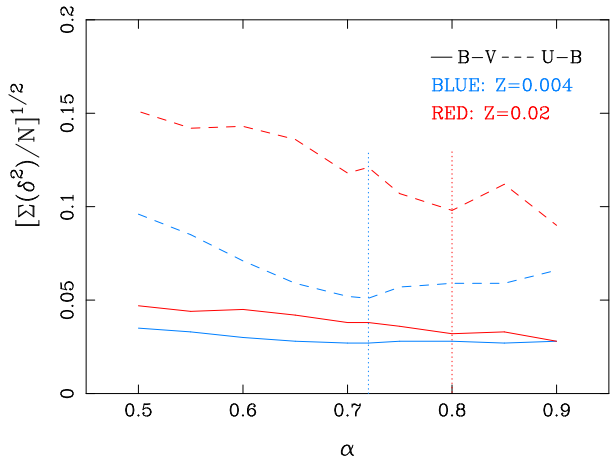}
\caption{The variance of ZAMS fitting to unevolved members in CM diagrams is 
plotted as a function of $\alpha$. Symbols and color distinguish the results 
of fitting in the CM diagrams $V,(U-B)$ (dashed lines) and $V,(B-V)$ 
(continuous lines). Results from fits to solar metallicity ZAMS (red lines) 
and ZAMS-Z004 (blue lines) are plotted. Vertical dotted lines indicate the 
minimum variance in both cases.}
\label{fita}
\end{figure}

We have performed a similar fit, but only to the ZAMS, and selecting stars 
that can be considered little evolved, constant reddening members. The sample 
consists of seven stars, five of them among the members by \cite{tur93}, and 
our four best candidates to non evolved MS members (actually two of them are 
in common with the \cite{tur93} members sample). Calculations are carried out 
for ZAMS of solar metallicity \citep{all82} and of Z=0.004 (ZAMS-Z004 
described above). Once a reference line and a reddening slope $\alpha$ are 
chosen, the absolute magnitude $M_V$, and color excesses $E(B-V)$ and 
$E(U-B)$ are determined. The average $V-M_V$ 
and color excesses are used to shift the ZAMS lines in the CM diagrams. For 
 each $\alpha$, we calculate the quantity $[\Sigma(\delta^{2})/N]^{1/2}$, 
where $\delta$ is the distance of every star to the shifted ZAMS in either CM
diagram and N is the number of stars. This quantity is plotted versus $\alpha$
in Figure~\ref{fita}. We see in this plot that a minimum is attained for 
$\alpha$=0.72 when ZAMS-Z004 are used, whereas for the fitting to solar 
metallicity ZAMS, the result of \cite{tur93} would be recovered.

The color excesses of the unevolved MS stars, obtained with $\alpha$=0.72 are 
then used to calculate $E(V-I)$ from the $(V-I)$,$(B-V)$ CC diagram, and the 
absorption coefficient $R_V=A_V/E(B-V)=2.4\times E(V-I)/E(B-V)$. The result,
$R_V$=3.10, coincides with the value characteristic for an average Galactic 
extinction \citep{car89}, and is then adopted, together with $\alpha$=0.72, to 
calculate color excesses and distances for the cluster stars.


\begin{figure}
\centering
\includegraphics[angle=-90,width=9cm]{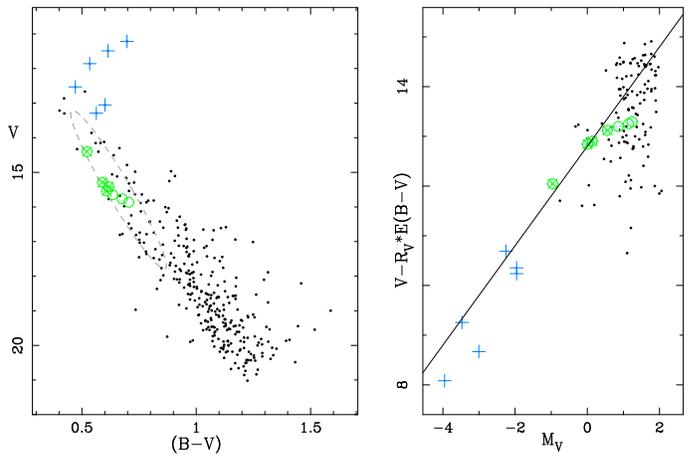}
\caption{Selection of upper MS member stars. The left panel shows 
all stars with measurement in all $UBVRI$ bands plotted in the upper 
$V,(B-V)$ CM diagram. A dashed line sketches the area where the 
non-evolved MS members are expected. Stars simultaneously located in 
similar areas in all four CM diagrams are plotted as circles. In the 
right panel, $V_0 \equiv V-R_V\times E(B-V)$ is plotted versus $M_V$ 
for stars brighter than $V$=18. A straight line represents the value 
DM=12.8. As in the left panel, circles represent stars considered as 
possible unevolved MS members. Among these, crossed circles are the 
four stars definitely selected as such, for calculation of cluster 
parameters. Crosses are the assumed post-MS members. }
\label{fig-cm}
\end{figure}

\subsection{MS Membership, Distance and Age}
\label{mem}

\subsubsection{Unevolved MS members}
\label{ms-mem}

To select reliable unevolved MS members we use a method that combines 
all color indices from $UBVRI$ photometry \citepalias{del07}. 
Figure~\ref{fig-cm} illustrates this selection. In the left panel we 
plot the upper part of the 
$V,(B-V)$ CM diagram, including all stars with measured values in all 
five $UBVRI$ bands. A dashed-line ellipse describes the region of the 
diagram where the non-evolved MS cluster members are expected, if any. 
The simultaneous location of stars inside similar regions in all four 
CM diagrams, together with their consistent location in the CC diagram, 
results in the sample, marked with circles in both panels of 
Fig.~\ref{fig-cm}.

In the right panel, we represent the quantities $V-R_V\times E(B-V)$ 
versus $M_V$, only for stars brighter than $V$=18, to make the plot 
clearer. A straight line for DM=12.8 is plotted as  an indication. In 
this 
right panel we observe the presence of foreground and background star groups, 
respectively, below and above the distance modulus value described by the 
selected unevolved MS members. Three of these appear to have a relatively 
lower distance modulus, and are further rejected. The final sample considered 
for the calculation of cluster distance modulus are those stars plotted as 
crossed circles. In both plots, crosses represent the stars adopted as 
post-MS members \citep{mof79,tur93}, to be considered in principle for the 
estimation of post-MS age.

\subsubsection{Color excess, distance, and post-MS age}
\label{distance}


\begin{figure*}
\centering
\includegraphics[angle=-90,width=16cm]{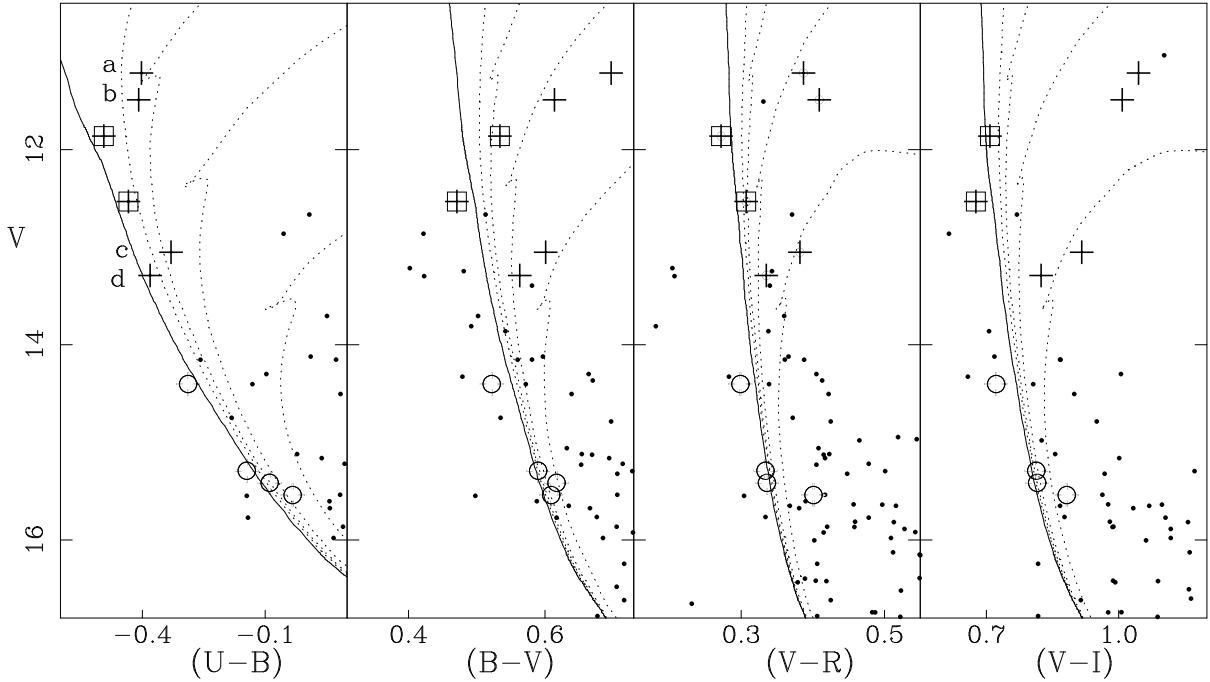}
\caption{Comparison of upper cluster sequence to isochrones in the 
CM diagrams. The ZAMS line for Z=0.004 (continuous line. See text) and 
Geneva isochrones for the same metallicity and LogAge(yr) 7.0, 7.2, 7.5 
and 8 (dotted lines, from left to right in each panel) are plotted. 
Circles are stars adopted as unevolved MS members. Crosses are stars 
adopted as post-MS members. Among them, crossed squares are those stars 
considered to estimate the age. The position of stars labeled $a-d$ is 
commented in the text. A quantitative comparison (see text) gives 
LogAge(yr)=6.51$\pm$0.07 for the post-MS cluster age.}
\label{fig-postMS}
\end{figure*}

The average values and rms deviations of color excess and distance 
modulus for the unevolved MS members amount to $E(B-V)$=0.78$\pm$0.02, 
DM=12.8$\pm$0.2. They are used as reference values to estimate membership for 
the remaining stars, following the procedure explained in detail in 
\citetalias{del07}. 

An estimate of the post-MS cluster age is obtained from the comparison to 
post-MS isochrones, where we use the set of models from the Geneva-isochrones 
for Z=0.004 and solar mixture. The comparison is illustrated in 
Fig.~\ref{fig-postMS}, where the upper CM diagrams $V,(U-B)$, $V,(B-V)$, 
$V,(V-R)$, and $V,(V-I)$ are plotted, with indication of the selected 
unevolved MS- and post-MS members (See. Fig.~2). To avoid the possible 
influence upon these diagrams of binarity \citep{gol74} or emission lines 
\citep{mer82}, we restrict our comparison to the two stars marked with 
crossed squares in the figure. A quantitative estimate, with extrapolation in 
 each diagram to the age values which would exactly reproduce the 
locations 
of both stars, gives 8 formal age values, one for each star in every one of 
the four CM diagrams. The average, and rms deviation of these values amount 
to LogAge(yr)=6.51$\pm$0.07, or 3.24$\pm$0.5 Myr, in good agreement with age 
values (3.8 Myr) for Dolidze~25 and its associated H{\sc II} region, quoted 
in \citetalias{pug09}. 

We note the significant difference between our distance result, and the
published distances for the cluster, around D=5.3~kpc \citep{mof75,tur93}, a 
difference which originates mainly in the use of low metallicity ZAMS in the
distance estimate. 

The effect of metallicity changes upon the cluster distance was discussed by 
\cite{tur93}, considering the expected variations due to a Galactocentric 
metallicity gradient. The key point here is however the low metallicity value 
obtained by \cite{len90}, Z=0.004, which does not follow the predictions of 
any gradient, as mentioned above.  
The question actually concerns absolute abundances rather than relative ones. 
We recall in this context the results by \cite{prza08}, who obtain a much
tighter results than previous works for B-type stars abundances in the solar 
neighbourhood.  
These results suggest that previous
 metallicity measurements are systematically too low, but the
 precise value of this offset and its application to stars located
 far outside the Solar radius is not clearly established. We
 therefore prefer not to adjust the metallicity value. We note
 that a "best guess" offset would change our distance estimate
 from 3.6 to 4.0 kpc and leave the rest of our conclusions essentially
 unchanged.

\subsubsection{PMS Membership and Age}
\label{pms-mem}

The remaining stars are now studied on the basis of the adopted 
values of distance and color excess. The method has been explained 
in detail in \citetalias{del07}. It consists in comparing color 
excesses and distance moduli of  each star with the average values 
obtained for the unevolved MS members. Distances are measured with 
respect to ZAMS and with respect to PMS isochrones from the Y2 set, 
for metallicity Z=0.004, and ages from 1 to 10 Myr. In this way,  each 
star has assigned several pairs of values of color excess and distance, 
and is considered either MS or PMS member, when one of these pairs agrees  
with those obtained for the unevolved MS members. Photometric uncertainties 
are considered in this assignement, and allowance is made for color excesses 
well above the average of the values for unevolved MS members (see 
\citetalias{del07}).


\begin{figure*}
\includegraphics[width=9cm]{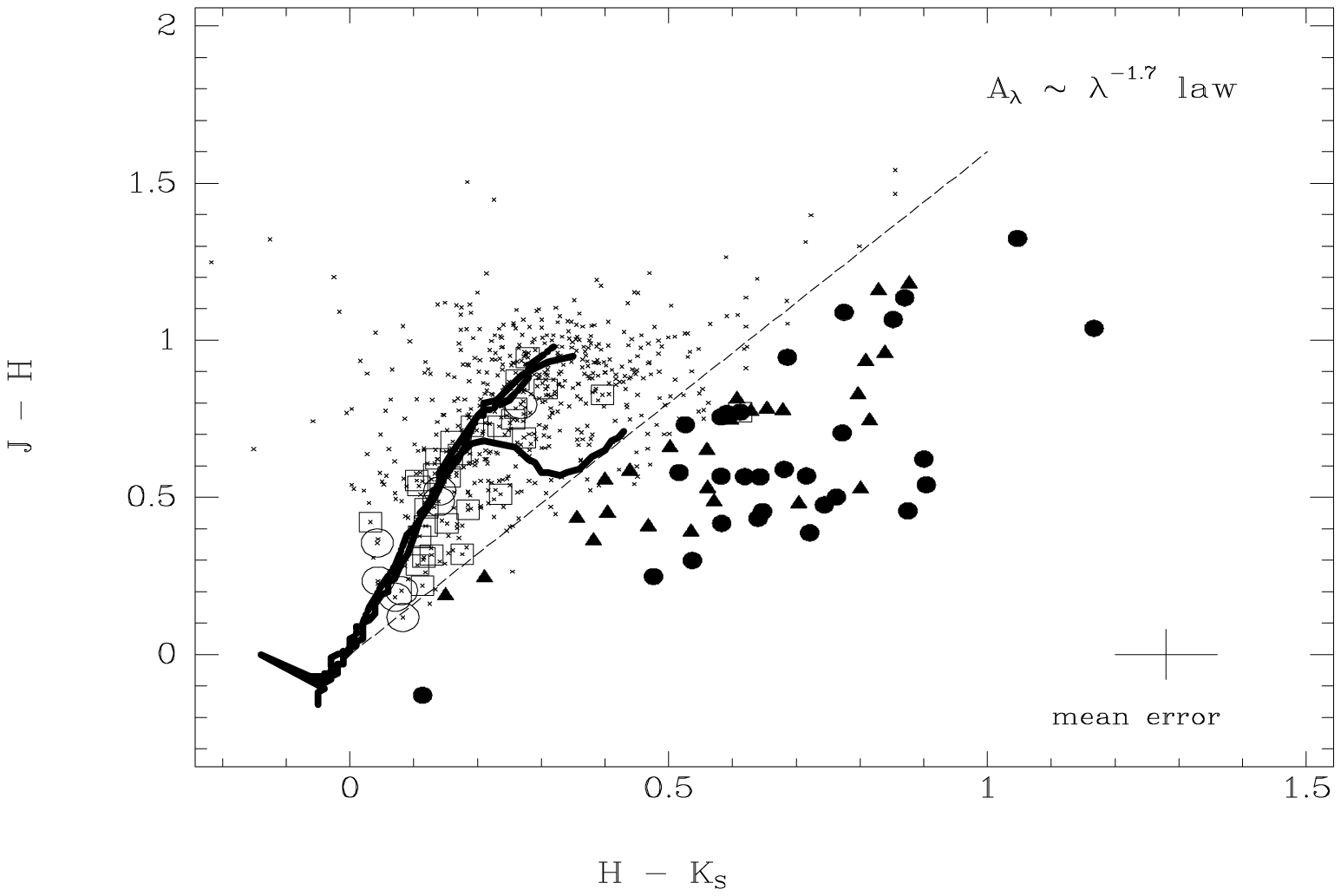}
\includegraphics[width=9cm]{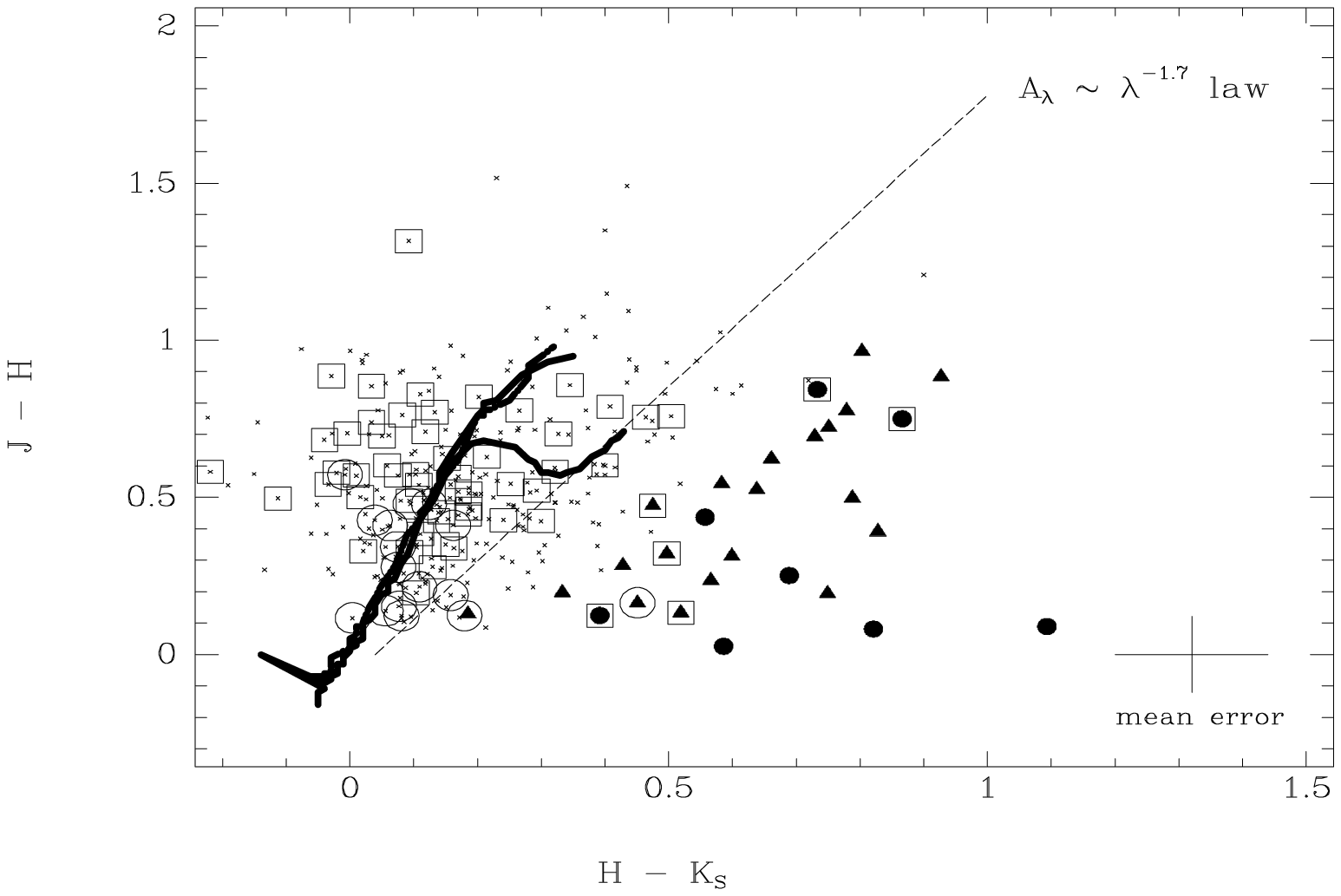}
\caption {The $J-H$/$H-K$ diagrams for the two near-IR datasets used.
{\bf Left:} The 3.5' $\times$ 3.5' central part of Do25 with deep
NOTCam imaging and a total of 651 sources. {\bf Right:} The 310 
2MASS point sources with ABCD flags found in the 6.5' $\times$ 6.5' 
field of our optical study. 
{\bf Both diagrams:} The bold curves outline the loci of main-sequence 
and giant stars. The reddening vector of an A0 star is shown as a 
dashed line. Bona-fide IR-excess sources (filled circles) are located 
more than 2$\sigma$ of the individual errors in $J-H$ and $H-K$ to the 
right and below the reddening line. Probable IR-excess sources (filled 
triangles) are those separated by between 1 and 2$\sigma$ only. The 
optically selected cluster members from Sect.~\ref{ubvri-results} are 
marked with large open circles (MS stars) and large open squares (PMS 
stars) for reference.
}
\label{fig-jhk}
\end{figure*}

From these comparisons, a star can be assigned as member with respect to 
several PMS isochrones, whose mean age value provides an age estimate for the 
PMS candidate. The age and uncertainty of the candidate PMS sequence 
is then obtained from the average value of the ages obtained for all 
candidates, and its rms deviation. This operation results in the value 
LogAge(yr)=6.7$\pm$0.2. The observed PMS sequence 
seems therefore to be coeval with the more massive MS members, in 
contrast with evidence found in other young clusters 
\citep[and references therein]{del04,del06,del07}, where both cases of 
PMS sequences younger and older than the corresponding MS members are 
found. 

%
%

\section{Results of the JHK$_S$ imaging}
\label{jhk-results}

\subsection{Near-IR excess sources}
\label{nirex}

\begin{figure}
\includegraphics[width=9cm]{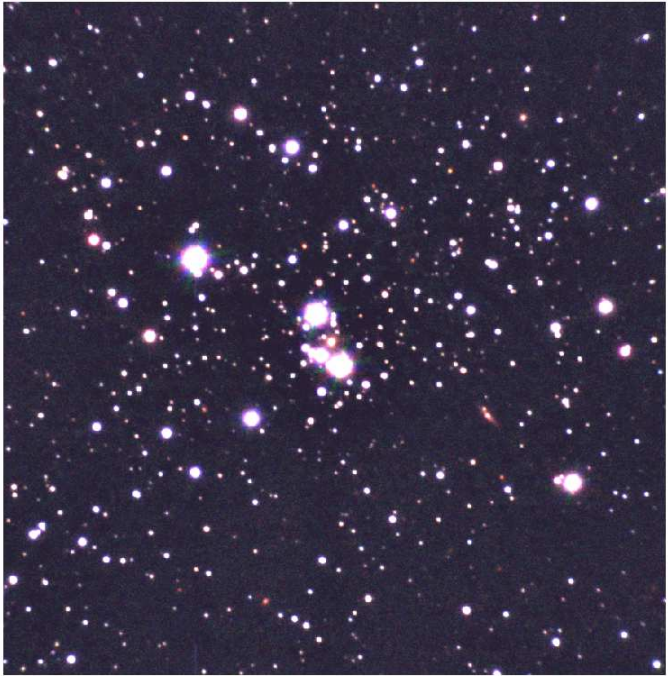}
\caption {RGB image of the NOTCam deep $JHK_S$ imaging of the 
3.5' $\times$ 3.5' central part of Dolidze~25. North up, east left. 
}
\label{fig-jhkimage}
\end{figure}

The $J-H/H-K$ diagram for 651 sources in the 3.5' field of deep JHK
imaging with NOTCam (see Fig.~\ref{fig-jhkimage}) is shown in the 
left panel of Fig.~\ref{fig-jhk}. 
The loci of main-sequence, giant and supergiant stars \citep{koo83} 
are indicated with bold curves. A reddening slope of 
E($J-H$)/E($H-K$) = 1.6 is calculated based on the NOTCam $JHK_S$ 
filter passbands (cf. Sect.~\ref{jhk-obs}) and using the 
$A_{\lambda} \propto \lambda^{-1.7}$ 
parametrization of the near-IR extinction law \citep{whi88}. The 
majority of the sources are clustered around
the loci of main-sequence and giant stars, and only marginally 
displaced along the reddening vector, in agreement with the average 
low extinction over this area as estimated in Sect.~\ref{reddening} 
($A_V$ = 2.4 mag, $A_K$ = 0.24 mag). 

Sources located to the right and below the reddening line in the
$J-H/H-K$ diagram - i.e. the above calculated reddening vector fixed
for an A0 star - have excess emission in the near infrared. In order 
to account for the uncertainties in the observed colours, we define 
as {\em bona-fide} IR-excess sources those located in this area and 
separated from the reddening vector by more than 2 $\sigma$ of the 
individual errors in the colour indices $J-H$ and $H-K$ (see 
Sect.~\ref{jhk-obs}). Those sources that are separated from the 
reddening vector by a distance between 1 and 2 $\sigma$ only are 
designated {\em probable} IR-excess sources. For the 3.5' $\times$ 
3.5' deep field we extract a sample of 29 bona-fide and 26 probable 
IR-excess sources. With this relatively conservative criterium we
will not sample all the IR-excess sources.
Because deep $JHK_S$ imaging was obtained in only part of the 
6.5' $\times$ 6.5' field studied in the optical (see 
Sect.~\ref{ubvri-results}), we have used the 2MASS point source 
catalogue to search for IR-excess sources. Using the same near-IR 
extinction law as above, the reddening slope  E($J-H$)/E($H-K$) 
becomes 1.78 because of the slightly different $JHK_S$ filter 
passbands for the 2MASS survey at $\lambda\lambda$ 1.24, 1.66, 
and 2.16 $\mu$m, respectively. 
The $J-H/H-K$ diagramme for the 310 sources with ABCD flags is shown 
in the right panel of Fig.~\ref{fig-jhk}. We have applied the same 
selection criterium for near-IR excess as above. For the 6.5' $\times$ 
6.5' field we find using 2MASS a total of 8 bona-fide and 20 probable 
IR-excess sources. In this diagram we also mark those cluster members 
found from optical photometry that have IR-excesses.
In both samples the fraction of IR-excess sources with respect to 
total number is about 8-9\%.

The 55 IR-excess sources found in the deep NOTCam field are in the
magnitude range 12.6 $<$ K $<$ 19.2 with an average at K = 17.1 mag,
and only 5 of these have 2MASS counterparts. Among these 5 only 2
have IR-excess according to the 2MASS dataset. The remaining 3 have
failed to be detected as IR-excess sources in the 2MASS dataset. It
is reasonable that the shallower 2MASS survey detected fewer excess
sources since the individual errors in the colours are larger.
Only 9 of the total of 28 2MASS IR-excess sources are located inside
the NOTCam deep field. These are in the magnitude range 12.6 $<$ K
$<$ 15.5 and are all optically visible sources. Surprisingly, it may
seem at first, that only two of them are confirmed to be IR-excess 
sources with the deeper NOTCam photometry. Looking in detail at the 
colour 
indices we find that for all the 7 ``non-confirmed'' sources, the 
$J-H$ and $H-K$ indices vary by 0.25 - 0.6 mag between the two epochs 
Nov-1999 for 2MASS data and Jan-2006 for NOTCam data. (For sources 
without IR-excesses the median of the absolute differences between 
the two datasets is $<$ 0.07 mag in the colour indices.) It is also 
worth mentioning that 3 of those 7 ``non-confirmed'' IR-excess sources 
are optically selected cluster members (see Sect.~\ref{ubvri-results}).

We consider it a sufficient condition for an IR-excess source to 
have measurable IR-excess in one of the epochs, and we conclude 
that the combination of NOTCam and 2MASS data yields a total of 
81 IR-excess sources (55 from NOTCam plus 28 from 2MASS 
minus 2 overlapping). 
Thus, the lack of overlap between the two sets of IR-excess sources
is explained in terms of sensitivity difference (i.e.
difference in error bars) and most importantly: intrinsic source 
variability. Variability is expected for young stars and is often 
used as a criterium of stellar youth; see e.g. \citet{kaa99} for a 
study of the efficiency of this criterion compared to that of 
IR-excess for membership assignment in a young embedded cluster. 


\subsection{Cluster membership of near-IR excess sources}

The excess emission in the near-IR is attributed to thermally 
radiating hot circumstellar dust. Determined by the temperature 
distribution of the dust particles, a superposition of blackbodies 
is produced, giving an excess flux at IR wavelengths compared to 
the spectral energy distribution (SED) of a naked star. 
Whether the dust is spatially distributed in a disk (typical for 
Classical T Tauri stars) or more spherically distributed (typical 
for protostars) is reflected in the shape of the SED towards the 
mid-IR. With only $UBVRIJHK$ photometry for this region we can not 
say much about the distribution of the dust, only its possible 
presence.  Because the typical number fraction of protostars with 
respect to PMS stars with disks is about 1/10 in young clusters, 
we statistically expect our population of IR-excess YSOs to be 
dominated by Class~II sources. We can not exclude, however, 
that some of our IR-excess sources may be protostars of Class~I 
type.


\begin{figure}
\centering
\includegraphics[angle=0,width=9cm]{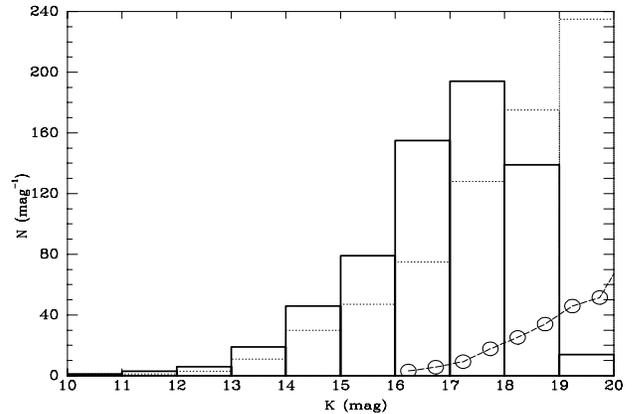}
\caption{K-band histogram of all sources in the 3.5' field (continuous 
histogram) compared to the histogram of K-band counts according to the 
Besan\c{c}on Galaxy model (dotted histogram). Circles show the number 
of extragalactic sources (Table~4 in \cite{bar09}) scaled to our field 
size and corrected for a cloud extinction of A$_K$ = 0.24 mag.
}
\label{fig-khist1}
\end{figure}

In Fig.~\ref{fig-khist1} we plot the $K$ band histogram of all our 
observed sources in the deep 3.5' $\times$ 3.5' field as well as that 
calculated from the Besan\c{c}on model of our Galaxy \citep{rob03}. 
In order to compare these numbers to the expected K-band counts of 
extragalactic origin, we have used the average values of the deep 
fields presented in Table~4 of \citet{bar09}.
Their numbers were scaled down to the size of our field and corrected 
for cloud extinction (A$_K$ = 0.24 mag) by shifting the values 
a quarter of a bin size to the right. From the figure it is 
evident that our observations are complete to around K = 17.5 mag. The 
level of extragalactic contamination in this bin is only $\sim$ 7\% 
and can not explain the excess of observed compared to model counts.
In the magnitude range 18 $< K <$ 19 mag, however, it is possible that 
$\sim$ 20\% of our source detections are distant galaxies. 

How the above percentages translate to percentages in the IR-excess 
sample is not obvious. 
Examining the map of cluster member candidates, however, we find that 
the faintest IR-excess sources are preferentially located in the central 
part of the image following the higher spatial density in general. Thus
from the spatial distribution it is unlikely that the IR-excess sample 
in the magnitude range 17 $< K < $ 19 consist of extragalactic sources, 
but we note that although we list our 81 IR-excess sources from 
Sect.~\ref{nirex} as YSO candidates, in the faintest bins there is a 
probability of extragalactic contamination possibly of the order of 
10-20 \%. 

\subsection{Comparison with optically selected cluster members}

A total of 104 pre-main-sequence (PMS) members were found in the 
6.5' $\times$ 6.5' area covered by our UBVRI photometry (see 
Sect.~\ref{pms-mem}). There are 2MASS counterparts with ABCD flags 
in all bands for 57 of these, of which only 5 have IR excess (see 
right panel of Fig.~\ref{fig-jhk}). Thus, less than 10\% of the 
optically selected PMS members that are also detected in $JHK$
with 2MASS, are found to exhibit excess emission in the near-IR. 
Inside the 3.5' $\times$ 3.5' small deep field there are 31
optically selected PMS stars of which only 3 have IR excess.

Only 19 of the 62 IR-excess sources are optically visible in the $V$ 
band - a necessary condition for our optical PMS selection. There are 
only 3 sources with both PMS {\em and} IR-excess classification in 
this field (these have ID numbers 468, 903, and 1563 in 
Table~\ref{tab-1}). Thus only 3/19 or 16\% of the IR-excess sources 
that are optically visible are classified as PMS 
members in our optical study. It is clear from the left panel of 
Fig.~\ref{fig-jhk} that practically all optically selected PMS stars 
are found quite near the loci of main-sequence stars in the $J-H/H-K$ 
diagramme and definitely to the left of the reddening line.


\begin{figure}
\centering
\includegraphics[angle=0,width=9cm]{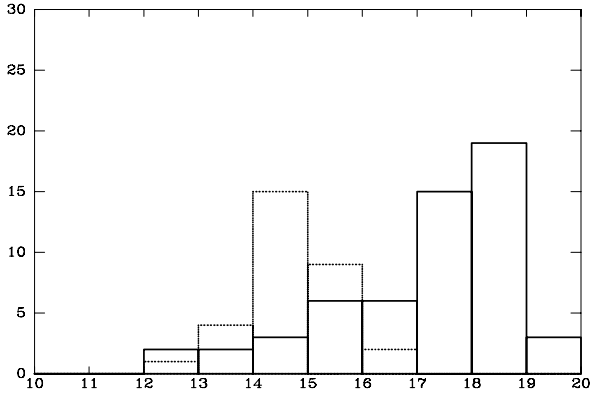}
\caption{{\bf Left:} K-band histogram of the two samples of young 
cluster member candidates: optically selected PMS stars (dotted line) and 
IR-excess sources (continous line).
}
\label{fig-khist2}
\end{figure}

The fact that most of the optically selected PMS stars are found
without signs of optically thick circumstellar disks, is either 
because they are so evolved (or independently of age they have lost
their circumstellar material), or the IR imprints of the disks are 
not detectable in the near-IR. 
It is well known from mid-IR studies \citep[e.g.][]{kaa04} that only 
about 50\% of the IR-excess sources in young embedded clusters show 
up with excesses in the $J-H/H-K$ diagramme. In the case of Dolidze~25, 
which is not really embedded, it is possible that only around 16\% of 
the optically visible PMS stars have disks, but this can only be assesed 
using mid-IR photometry.

Using two different photometric methods we have two samples of PMS
cluster members with very little overlap. The optically selcted
PMS stars have a magnitude range of 12.6 $<$ K $<$ 16.6, i.e. all 
being relatively bright in the near-IR. The near-IR excess sources 
on the other hand span the larger magnitude range 12.6 $<$ K $<$ 19.2
with a median at K = 17.8 mag. Thus, in general they comprise a 
fainter population, see Fig.~\ref{fig-khist2}. Correcting for average
extinction (A$_K$ = 0.24 mag) and distance (DM = 12.8 mag) we get
absolute magnitudes from -0.4 $<$ M$_K$ $<$ 6.2 mag for the IR-excess
sources. Assuming a typical age of 2-3 Myr for the IR-excess sources, 
the sub-stellar limit is around M$_K$ = 5.5 mag according to PMS 
evolutionary tracks \citep{bar98}. The mass range of the IR-excess 
population extends thus to well beyond the brown dwarf limit, while
our completeness estimate at $\sim$ K=17.5 mag corresponds to 
M$_K$=4.5 mag only.

This statistical result shows that the optical photometric method and
the near-IR excess method sample different PMS populations, possibly 
with an age difference, but not necessarily so. The multi-wavelength 
approach covering both optical and near-IR bands in photometric studies 
of clusters is advantageous because the two methods are complimentary. 
This same conclusion was drawn from a similar type of study of the 
double cluster in Sh2-294 \citep{yun08}. 

%
%

\section{Discussion}
\label{disc}

\subsection{Characteristics of the Dolidze 25 cluster}

A deep UBVRIJHK photometric study of the Dolidze 25 cluster, in a
6.5' field centred around the B1 stars TYC~148-2558-1 and 
TYC~148-2577-1, SIMBAD names for our ID 1060 and 1362, i.e. WEBDA 
17 and 15 \citep{mof79}, is presented in the previous sections. This 
field covers an area of almost 7~$\times$~7 pc, using our new 
distance estimate of 3.6~kpc described in Sect.~\ref{distance}. The 
analysis of cluster membership resulted in a total of 214 candidate 
cluster members: 35 main-sequence (MS) and post-MS members, 104 
optically selected pre-main sequence (PMS) members, and 81 IR-excess 
sources, probably young PMS stars with circumstellar disks (Class II 
sources), although Class I sources can not be excluded. Six of the 
IR-excess sources coincide with optically selected PMS stars.


\begin{figure}
\centering
\includegraphics[width=9cm,angle=-90]{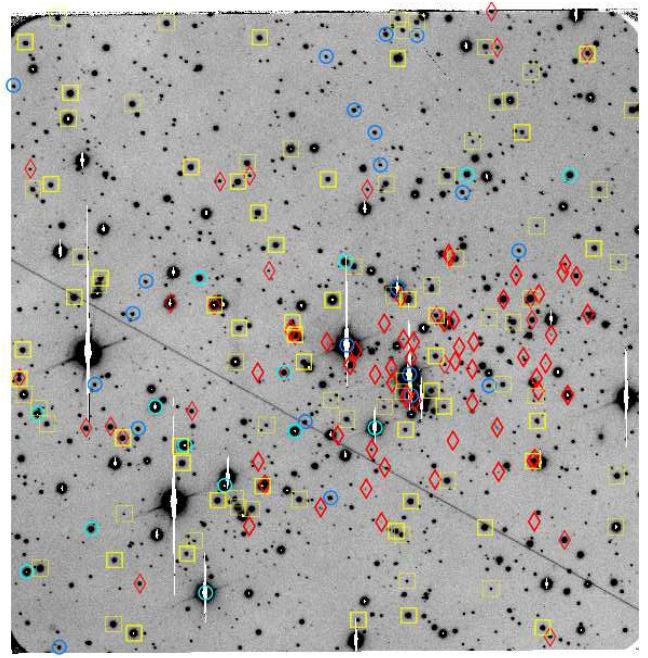}
\caption {ALFOSC R-band image of the 6.5' $\times$ 6.5' area (north up
and east left). The location of MS and post-MS stars of the younger
generation (cyan circles) and of the older generation (blue circles), 
optical PMS candidates (yellow squares), and IR-excess sources (red diamonds) 
are shown. Note that the IR-excess source selection has a higher 
sensitivity in the smaller 3.5' $\times$ 3.5' area covered by NOTCam.
}
\label{fig-alfoscfig}
\end{figure}

\subsection{Age estimates}
\label{age}

\begin{figure}
\centering
\includegraphics[angle=-90,width=9cm]{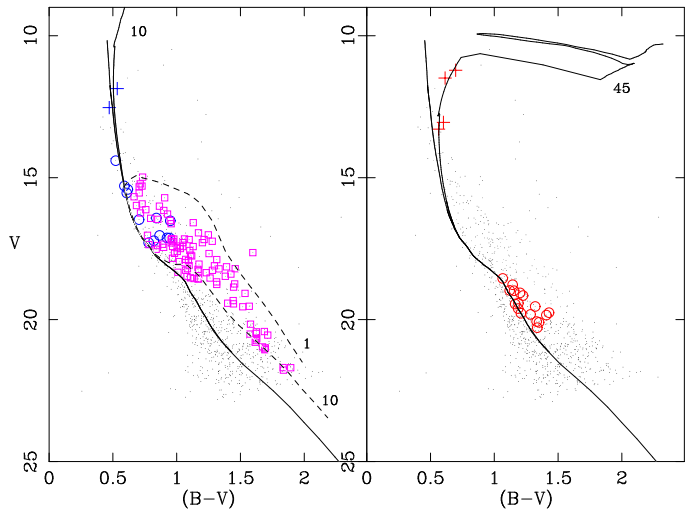}
\caption{$V,(B-V)$ diagrams with indications in the cluster members assigned 
to two generations of stars at different ages. In both panels, all stars are 
represented with small dots. In the left panel, proposed young members 
(3-5 Myr) are plotted, with post-MS (crosses) and MS members (circles), 
and PMS members asigned in at least 3 CM diagrams (squares). Cluster 
members in the older generation are plotted in the right panel, with the 
same meaning for the symbols. ZAMS-Z004 line (continuous thick), 10 Myr 
isochrone (continuous thin), and PMS isochrones of ages 1 and 10 Myr 
(broken) are plotted and labeled in the left panel. The same ZAMS, and 
the 45 Myr isochrone are plotted in the right panel. }
\label{fig-twogen}
\end{figure}

The membership analysis of the central Dolidze~25 cluster shows a 
young generation of stars, of around 3-5 Myr, with a rich population 
of PMS candidate members, coeval to the upper massive MS stars. 
However, this analysis also reveals the presence of older stars, 
which are selected as MS members of lower mass and luminosity than 
the selected PMS candidates. This suggests the presence in the cluster 
of two generations of stars. In this hypothesis, the four member stars 
labeled with $a$ to $d$ in the left panel of Fig.~\ref{fig-postMS}, can 
be interpreted as the evolved upper sequence of the older star 
generation. To illustrate this point we show a comparison in 
Fig.~\ref{fig-twogen}, where the $V,(B-V)$ diagrams of these suggested 
generations are plotted separately.


\begin{table}
\caption{The 35 MS and post-MS members found in our analysis of a 6.5'
wide field. The last column indicates the age group, where old means
around 40 Myr and young 3-5 Myr. For fluxes, colours and coordinates we 
refer to Table.~\ref{tab-1} (available on-line only).}
\newcommand\cola {\null}
\newcommand\colb {&}
\newcommand\colc {&}
\newcommand\colf {&}
\newcommand\colg {&}
\newcommand\eol{\\}
\newcommand\extline{&&&&&\eol}
\begin{tabular}{lllccrr}
\hline
\hline
\cola ID\colb WEBDA\colc 2MASS\colf V\colg Age group\eol
\cola \colb \colc \colf (mag)\colg \eol
\hline
\cola  298\colb    \colc 06445568+0015307\colf 17.034\colg young\eol
\cola  513\colb    \colc                 \colf 19.839\colg old\eol
\cola  649\colb    \colc                 \colf 19.756\colg old\eol
\cola  761\colb    \colc 06445984+0015311\colf 17.128\colg young\eol
\cola  786\colb    \colc                 \colf 19.152\colg old\eol
\cola 1022\colb    \colc                 \colf 19.775\colg old\eol
\cola 1053\colb   16\colc 06450210+0013165\colf 13.052\colg old\eol
\cola 1060\colb   15\colc 06450217+0013294\colf 11.490\colg old\eol
\cola 1121\colb   18\colc 06450266+0014218\colf 13.290\colg old\eol
\cola 1179\colb    \colc                  \colf 18.549\colg old\eol
\cola 1214\colb    \colc                  \colf 20.284\colg old\eol
\cola 1243\colb   19\colc 06450356+0012569\colf 12.532\colg young\eol
\cola 1244\colb    \colc                  \colf 19.611\colg old\eol
\cola 1329\colb    \colc                  \colf 19.058\colg old\eol
\cola 1362\colb   17\colc 06450471+0013472\colf 11.215\colg old\eol
\cola 1371\colb  197\colc 06450477+0014379\colf 16.479\colg young\eol
\cola 1422\colb  196\colc                 \colf 18.772\colg old\eol
\cola 1441\colb    \colc                  \colf 20.068\colg old\eol
\cola 1522\colb    \colc                  \colf 18.971\colg old\eol
\cola 1560\colb  123\colc 06450680+0012549\colf 15.539\colg young\eol
\cola 1609\colb  119\colc 06450726+0013306\colf 17.292\colg young\eol
\cola 1863\colb  110\colc 06450966+0012217\colf 15.291\colg young\eol
\cola 1930\colb   22\colc 06451043+0011164\colf 11.861\colg young\eol
\cola 1943\colb  174\colc 06451066+0014281\colf 17.221\colg young\eol
\cola 2003\colb  107\colc 06451127+0012463\colf 15.414\colg young\eol
\cola 2095\colb  102\colc 06451246+0013096\colf 14.402\colg young\eol
\cola 2130\colb     \colc                  \colf 19.816\colg old\eol
\cola 2157\colb     \colc                  \colf 18.982\colg old\eol
\cola 2188\colb     \colc                  \colf 19.422\colg old\eol
\cola 2289\colb     \colc                  \colf 19.450\colg old\eol
\cola 2308\colb     \colc 06451507+0011553\colf 17.121\colg young\eol
\cola 2423\colb     \colc 06451636+0010433\colf 19.538\colg old\eol
\cola 2499\colb     \colc 06451724+0013054\colf 16.519\colg young\eol
\cola 2533\colb     \colc 06451767+0011292\colf 16.421\colg young\eol
\cola 2581\colb     \colc                 \colf 20.097\colg old\eol
\hline
\hline
\end{tabular}
\label{tab-2}
\end{table}

The main indication that supports this suggestion is the presence of low mass 
MS candidate members, together with the very good visual fit of the older 
isochrone to the redder post-MS stars. A formal age calculation for the
older post-MS stars in the right panel of Fig.~\ref{fig-twogen}, performed in 
the way explained in Sect.~\ref{distance}, gives LogAge(yr)=7.6$\pm$0.2. On 
the other hand, MS candidate members in the $V$ range 16-18 mag could belong 
to both generations, and some of them could indeed be PMS stars approaching 
the MS. Even some of the faint MS members ($V >$ 18), namely those with redder 
$(B-V)$ index (Fig.~\ref{fig-twogen}, right panel), could also be the 
last signs of a PMS sequence in the older star generation. The precise 
status of each star cannot be ascertained here, since all candidate 
members are at the same distance, and with comparable color excess: 
these are actually the criteria to propose them as cluster members. A 
detailed spectroscopic study, aiming at the simultaneous determination 
of spectral type and radial velocity for these stars \citep{del99,del04} 
is planned as the appropiate tool to confirm the suggested age structure.

In this context we remark that the indications of age spread, and 
sequential or continued star formation in young clusters are widely 
reported in the literature 
\citep[i.e.][and their references]{sub06,bha07,del07}. Evidence of 
episodic, and sequentially triggered 
star formation, has been suggested in large forming regions of the 
Small Magellanic Cloud \citep{sab07,car07}, and NGC6946 \citep{lar02, san09}. 
The interesting point in the 
present case, if confirmed, is the presence of two star generations in 
the same place. 

The interpretation of the stars labeled $a-d$ in Fig.~\ref{fig-postMS} as 
evolved cluster members requires some comments. Their location 
in the $V,(V-R)$ and $V,(V-I)$ CM diagrams supports this hypothesis, which 
is however contradicted by the position of the stars in the $V,(U-B)$ CM 
diagram (see Fig.~\ref{fig-postMS}). In this last diagram, the comparison 
to isochrones under the assumption that they are evolved cluster members, 
suggests an age of at most 15 Myr. One possibility to solve the discrepancy 
is to assume that they indeed are as young as indicated by their $(U-B)$ 
color, while the other three color indices would just reveal, either simple 
reddening, with a peculiar reddening slope, or some color excess, which 
increases towards redder indices. 

In the line of considering these stars younger than proposed here, we might 
recall recent results on massive young clusters, where the presence of a 
noticeable supergiant population is argued \citep{cla09,ale09}. The 
assumption that the brightest stars in our CM could indicate such a 
population would result in a distance from them around 7~kpc, which 
would discard them as cluster members, even under consideration of the largest 
distance estimates given for Dolidze~25 in the literature. 

On the other hand, the presence in these stars of color excesses which 
are larger for redder colors is also uncertain. It is in any case 
contradicted by our NIR results, in which none of the proposed post-MS 
candidates appear to exhibit NIR excess (see Sect.~\ref{jhk-results}). 
This lack of 
NIR excess also weakens the suggestion of a redder binary companion 
which would show up in redder color indices. The performance of 
adequate spectroscopic observations is needed to elucidate among these 
possibilities, including the one put forward here, which considers the 
presence of two generations of stars in the cluster.

\subsection{Spatial distributions}
\label{spatial}


\begin{figure}
\includegraphics[width=9cm]{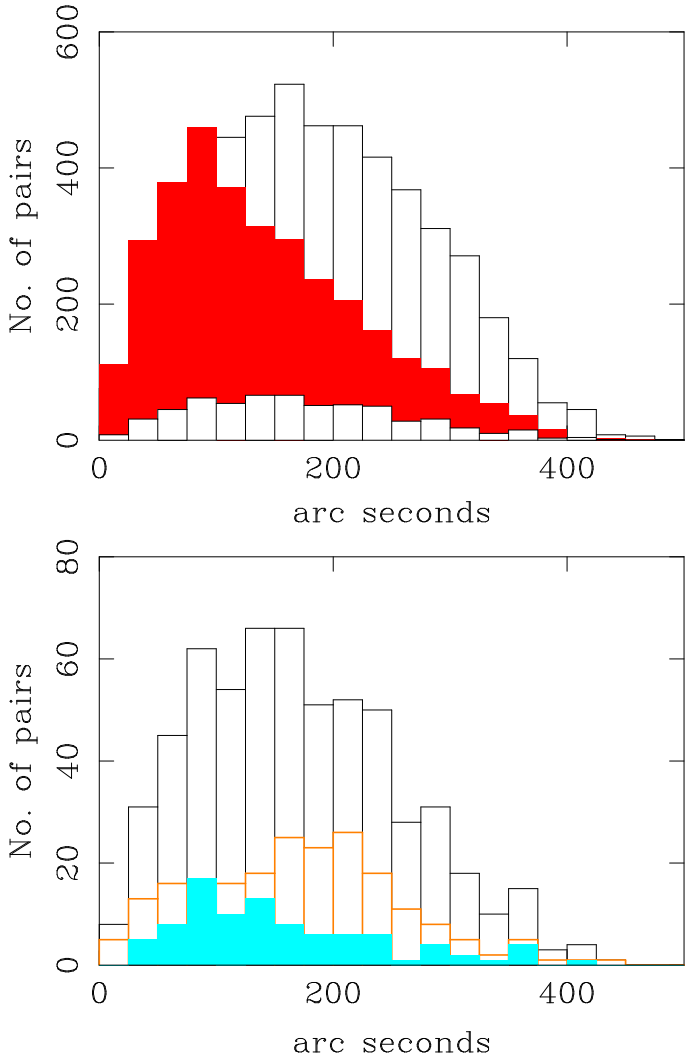}
\caption {The number of pairs as a function of separation (in arc
seconds), a measurement of the clustering properties of a population, 
is shown in histograms. {\bf Top:} 
The optically selected PMS stars comprise the large population in the 
white/open histogram, the IR-excess sources are shown as red/filled 
histogram, and the total population of MS and post-MS stars in the 
small white/open histogram. {\bf Bottom:} The total MS and post-MS 
population is shown in the white/open histogram, the older population 
in the orange/bold line histogram, and the younger population in the 
blue/filled histogram.
}
\label{fig-distribution}
\end{figure}

The spatial distribution of the various cluster members is shown in 
Fig.~\ref{fig-alfoscfig}, overlaid on an R-band image, the different 
colours and symbols distinguishing different populations. The 
MS and post-MS members have a clear zone of avoidance towards the 
south-west, otherwise they are few and relatively scattered. The 
IR-excess members seem to have a preference along a SE-NW diagonal 
belt with an enhanced surface density around ID 1362, which appears 
to be at the approximate cluster center (see also 
Fig.~\ref{fig-jhkimage}, the area over which we have deep $JHK_S$ 
imaging). There is an enhanced number density of sources both in the 
optical and near-IR images along the SE-NW diagonal belt. There seem
to be no apparent spatial preference among the optically selected 
PMS stars. 

We calculate the distribution of separations of pairs of members 
within the different sub-groups, a way to quantitatively assess 
possible clustering properties \citep{kaa04}, although projected in
the plane of the sky. In 
Fig.~\ref{fig-distribution} we show in histograms the number of 
pairs versus separation (in arc seconds). Strong peaks in this 
distribution indicate the scale of clustering or sub-clustering of 
the population, or more correctly: the approximate size/diameter 
of the densest region. The upper panel shows the distribution of 
pairs of optically selected PMS stars (largest group), IR-excess 
sources (medium group), and all MS and post-MS sources (small group). 
For the optically selected PMS stars there is a small peak around 
160'', translating to about 3~pc using our distance estimate of 
3.6~kpc, but the distribution is broad and indicates that the 
optically selected PMS stars are in general quite scattered. The 
IR-excess sources, on the other hand, have a much more pronounced 
peak, located at 90'' or 1.6~pc. We {|bf note} that the deep near-IR 
study was made in a smaller field, and only the shallower 2MASS 
survey could be used for the whole area, such that the histogram
should not be considered beyond 210'' for this group. There seems
to be a clear tendency of clustering of this population. 

The MS and post-MS population, which is a comparatively small group, 
is seen to have a relatively flat distribution in the upper panel. 
When separating the total MS and post-MS population into the two age 
groups suggested in the previous section (approximately 5 Myr and 
40 Myr), we see that the spatial distribution of the two is distinct. 
The younger generation (blue/filled histogram) is slightly more 
clustered, i.e. the distribution peaks at a smaller scale, than the 
older generation (orange/bold line histogram). The young generation
of MS and post-MS stars peaks around the same scale as the IR-excess
population.

Although the number statistics is relatively small, this gives an 
independent support to the hypothesis of two different generations 
of MS and post-MS stars suggested by isochrone fitting and described 
in Sect.~\ref{age}. 
The spatial distribution of a cluster population is expected to 
become more dispersed with time. Assuming a typical stellar velocity 
dispersion 
of $< \approx$ 1.5 kms$^{-1}$ found for old ($>$ 10$^8$ Myr) open 
clusters \citep{loh72}, one might expect 40 Myr old stars to have 
moved quite some distance away from their birth place. We also note 
that much larger velocity dispersions have been found for the young 
clusters NGC2244 and NGC6530 of 35 and 8 kms$^{-1}$, respectively 
\citep{che07}. Thus, part of the older population could be located 
outside the 7 pc wide field studied by us, but only a proper motion 
study will reveal the velocity dispersion in Dolidze~25. 

\subsection{Mass function}
\label{mass}

The quantitative measurement of ages described in Sect.~\ref{mem} above, 
provides mass estimates for the assigned cluster members, both from the 
ZAMS and post-MS isocchrones, and from the PMS isochrones as well. These 
mass values are averages of the values read from the corresponding 
reference lines, and are calculated in the same way as the age estimates. 

\begin{figure}
\includegraphics[angle=-90,width=9cm]{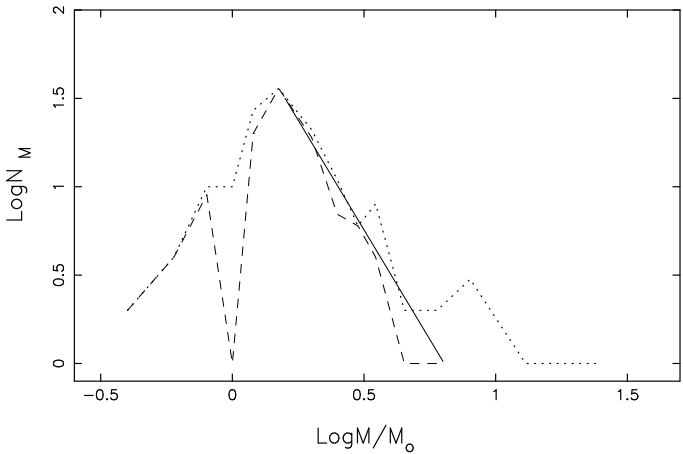}
\caption {Mass function for the candidate members in Dolidze 25. The 
models of \citet{lej01} and \citet{yi01} for Z=0.004 have been used to 
estimate membership and calculate masses for the assigned members. Mass 
functions of PMS candidates (dashed curve) and all members (dotted curve) 
are plotted. The continuous straight line reproduces the slope of the 
Salpeter IMF (-2.35), which is seen to fit very well the calculated mass 
function for PMS candidates of masses above LogM/M$_\odot$=0.2. 
}
\label{fig-massfun}
\end{figure}

We apply here these calculated mass values to illustrate the properties 
of the cluster mass function. In Fig.~\ref{fig-massfun} we plot the 
logarithm of the number of stars of mass value versus the logarithm 
of the star mass, in units of solar masses. The break down of the function 
at masses below 1.3 solar masses is probably due to an underabundance of 
low mass stars, but other factors, such as the incompleteness of the 
detected sample should be present, and we do not discuss further this 
feature. On the other hand, the over abundance of massive stars is probably 
due to the consideration of all MS and post-MS stars in this function, 
without distinction of the argued presence of two different star 
generations. In fact, no differences in mass function depending on 
metallicity are expected at these high masses \citep{bat05}.

The slope at masses above 1 M$_{\odot}$ should otherwise be more reliable, 
in view of the color excess and distance derived for the cluster, and the 
expected brightness range covered by these stars, which allow one to expect 
a higher degree of completeness in the cluster members detection. In this 
figure we note the present sequence of assigned candidate PMS members, 
with a mass function very well reproduced by a Salpeter mass function 
slope (-2.35 in the plot of the figure). 


\subsection{Comments on the age distribution}

There are some cases in the recent literature of regions similar to 
this, where a central, ageing cluster has caused a set of structures 
around it, driving in some cases the formation of new stars 
\citep{yun08,deh03}. Here we could be witnessing a region similar to 
these, but some 10 million years later than those other regions. 

In view of the possible presence of two generations of MS and post-MS 
stars in the central field, as suggested by our data in combination 
with isochrone fitting, we speculate that the bubble may have been 
shaped by the first generation of stars some 40 Myr ago. This 
population may be quite dispersed by now. The new generation, with 
ages approximately around 3-5 Myr found both for MS, post-MS, and PMS 
stars, may have been triggered by the first population, and although 
we see a cluster projected close to the centre of the bubble, its 
exact location may just as well be closer to the shell of the bubble 
in the line of sight towards us. As seen in Fig.~\ref{fig-UKST} the red 
structures without bright rims at the edges, are most likely cloud filaments 
and structure located on the near side of the bubble, and would be showing 
absorption relative to the H$\alpha$ emission background. These structures 
contrast to the ones with bright rims, i.e. enhanced H$\alpha$ emission at 
the edges. 

In this context, we may interpret the collection of Class~I sources in 
the very direction of the cluster core. As suggested by \citetalias{pug09}, 
they could be indeed embedded in the above mentioned H$\alpha$ absorbing 
structure, and be a part of the surrounding region, where other star forming 
centers are, seen only in foreground.  

All these indications enhance the view of S284 as a region with past events 
of star formation at different ages, and presently ongoing formation activity 
of new generations of stars, ranging in age from a few Myr down to very young 
objects in the first formation phases.

\section{Summary and Conclusions}
\label{sum}

\begin{itemize}

\item We obtained deep UBVRIJHK photometry in the central 6.5' part of the 
Dolidze~25 cluster, reaching the 10$\sigma$ limiting magnitudes of $V$ = 23.3, 
$J$ = 19.7, and $K_S$ = 18.6. Magnitudes, colours, and positions of all 
sources, as well as possible membership assignments, are published in 
Table~\ref{tab-1}, available on-line. 

\item The membership analysis revealed 214 candidate members, of 
which 35 are main-sequence or post-MS stars, and the rest are PMS 
candidates. The PMS candidates comprise optically selected sources 
as well as sources with IR excess. 

\item The use of low metallicity ZAMS and isochrones to estimate distance, 
gives a new distance for the cluster to be 3.6~kpc. 

\item The ages of the optically selected PMS stars are likely around
5 Myr and those of the IR-excess sources possibly slightly younger.

\item We find a possible existence of two generations of member stars,
one with MS and PMS stars of age below 5 Myr, and another older population 
of age around 40 Myr. The spatial distribution of the two potential 
generations is found to be distinct, giving independent support to this 
suggestion.

\item Dolidze~25 seems to be located in the centre of an ionized bubble 
which probably originated from the older generation of stars. We cannot 
exclude that the young cluster generation is located in the near side of 
the bubble shell, though.

\end{itemize}

In the previous presentation of results we have described 
the presence of post-MS cluster members at similar luminosity but 
different evolutionary states, as inferred from their color indices. 
At the same time, we detect the presence of MS candidate members with 
masses below the presently PMS stars. This evidence adds to the 
indication of the presence in the field of a wide range in distances, 
with foreground and background groupings of stars (see Fig.~\ref{fig-cm}) 
producing the general view of a region where star formation has been 
going on for a long time, both at different places of the H{\sc ii} 
region, and in the same place at different epochs. At present, we observe 
this mixture of populations, with a star forming process which takes 
place in an already formed cluster, giving rise to the presence of two 
star generations of different age and being members of the same cluster. 
The evolved state of some of the most luminous cluster members, together 
with the presence of low mass MS member stars, can indeed be interpreted 
in this context. We can not exclude, though, that the younger population 
is seen in projection in front of the older one, possibly being formed in 
the shell of the bubble. Finally, we recall that the interpretation of two
generations of member stars is furthermore favoured from inspection of the 
spatial distribution.

\begin{acknowledgements}
The NOT staff atronomers Tapio Pursimo and John Telting are warmly acknowledged for the ALFOSC observations carried out during service/technical nights. Thanks to the referee, A. Moffat, for his comments and recommendations which resulted in an improvement of the article. We also wish to tank D. Lennon for his very valuable and useful comments. The data presented here have been taken using ALFOSC, which is owned by the Instituto de Astrofisica de Andalucia (IAA) and operated at the Nordic Optical Telescope under agreement between IAA and the NBIfAFG of the Astronomical Observatory of Copenhagen. This work has been supported by the Spanish MICCIN through grants  AYA2007-64052, and by the Consejer\'\i a~ de Educaci\'on y Ciencia de la Junta de Andaluc\'\i a, through TIC 101 and TIC 4075 grants. EJA acknowledges the financial support from the Spanish MICINN under the Consolider-Ingenio 2010 Program grant CSD2006-00070: First Science  with the GTC. This research has made use of the SIMBAD database, operated at CDS, Strasbourg, France, the NASA ADS Abstract Service, and the WEBDA data base, developed by Jean-Claude Mermilliod at the Laboratory of Astrophysics of the EPFL (Switzerland), and further developed and maintained by Ernst Paunzen at the Institute of Astronomy of the University of Vienna (Austria). This publication makes use of data products from the Two Micron All Sky Survey, which is a joint project of the University of Massachusetts and the Infrared Processing and Analysis Center/California Institute of Technology, funded by the National Aeronautics and Space Administration and the National Science Foundation. We have also made use of the public data in the AAO/UKST H$\alpha$ survey. 
\end{acknowledgements}

\bibliographystyle{aa}
\bibliography{do25}

\end{document}